\documentclass[11pt]{article}
\usepackage{moriond,epsfig}

\def\be{\begin{equation}}
\def\ee{\end{equation}}
\def\bea{\begin{eqnarray}}
\def\eea{\end{eqnarray}}

\begin{document}
\vspace*{4cm}
\title{PHOTON+JET PRODUCTION AT $\sqrt{s}$=1.96~TeV}

\author{ C. DELUCA }

\address{Institut de F{\'i}sica d'Altes Energies, Universitat Auton{\`o}ma de Barcelona, \\
Ed. Cn, 08193 Barcelona, Spain.}

\maketitle\abstracts{
Prompt photon production results by the CDF and D\O~Collaborations in the Tevatron Run~II at a
center of mass energy of $\sqrt{s}$=1.96~TeV are presented. Cross sections for central isolated photons,
photon+jet production and photons produced in association with a heavy flavor quark are reported. The
measurements are compared to Next-to-Leading order perturbative QCD predictions.
}

\section{Introduction}
\label{sec:intro}

The prompt photon cross section is a classic measurement to test perturbative QCD (pQCD) with the
potential to provide information on the parton distribution function (PDF), and sensitive to
the presence of new physics at large photon transverse momentum. Prompt
photons also constitute an irreducible background for important searches such as 
H$\rightarrow\gamma\gamma$, SUSY or models with extra-dimensions with energetic photons in the
final state. From an experimental point of view, the study of direct photon production has several advantages
compared to QCD studies using jets, but require a good understanding of the
background, mainly dominated by light mesons ($\pi^0$ and $\eta$) which decay
into photons. Since these photons are produced within a jet, they can be suppressed by requiring the
photon candidates to be isolated in the calorimeter. 

\section{Inclusive isolated prompt photon cross section}
\label{sec:incpho}
\begin{figure}[tbp]
   \begin{center}
   \begin{tabular}{cc}
        \renewcommand{\tabcolsep}{0.01in}
\includegraphics[width=3.0in, angle=0]{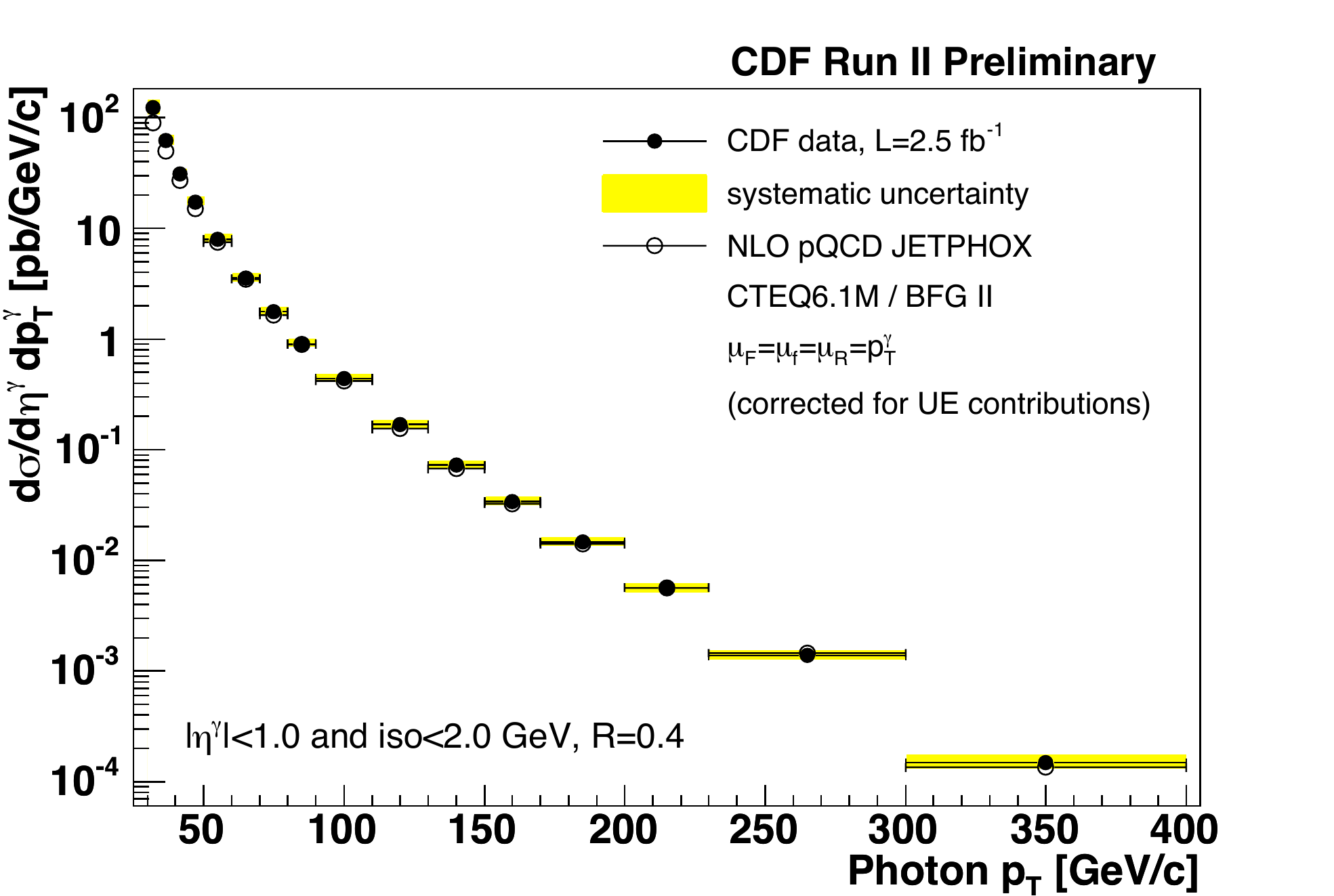} & \includegraphics[width=3.0in, angle=0]{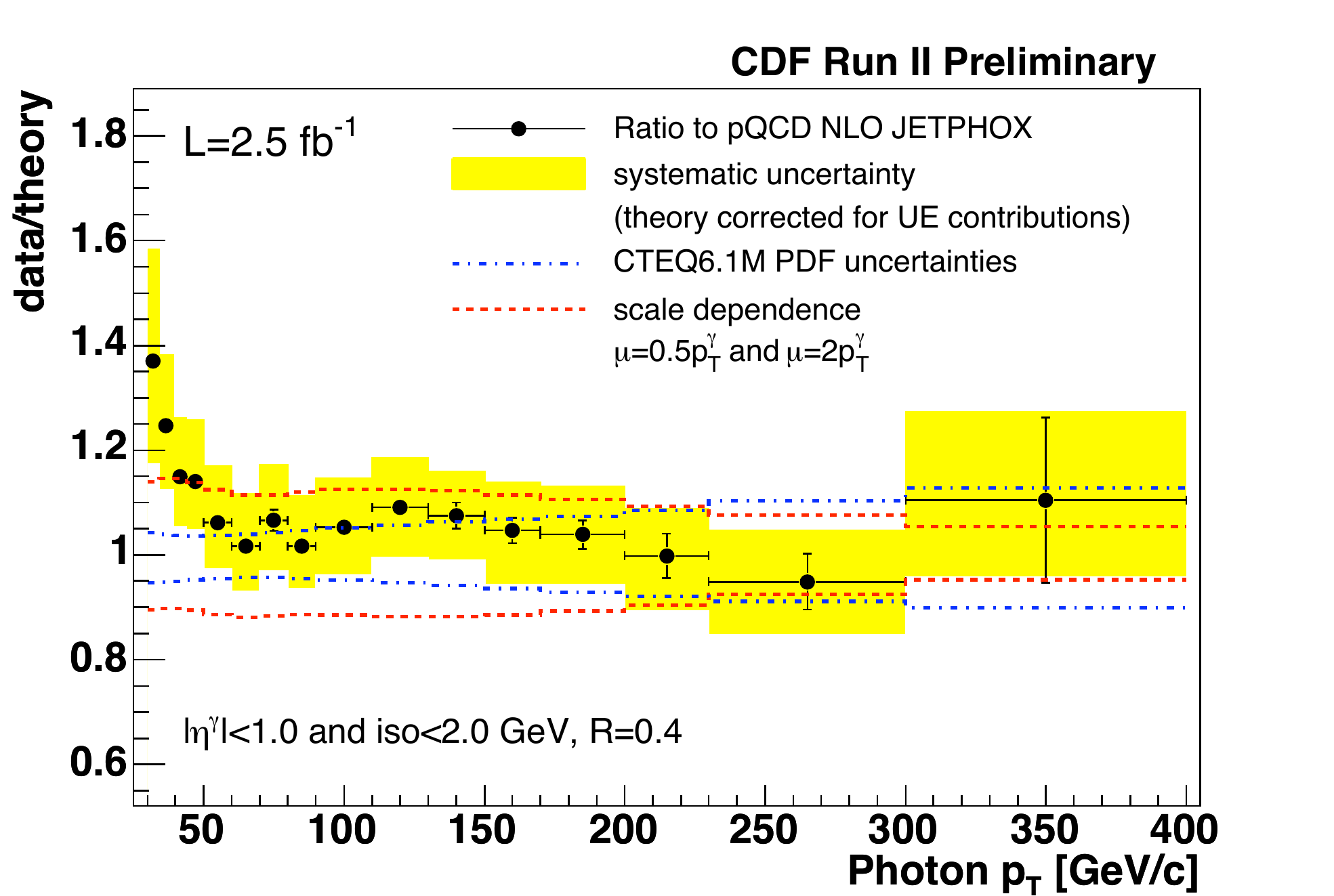} \\
   \end{tabular}
  \end{center}
     \caption{\small The inclusive isolated prompt photon cross section as a function of the photon
transverse momentum measured by the CDF experiment (left). The ratio data to the NLO pQCD predictions
corrected for the non-pQCD contributions of the underlying event is shown in the right. 
   \label{fig:incxsec}}
 \end{figure}

CDF has recently measured the inclusive isolated prompt photon cross section for central photons
using 2.5~fb$^{-1}$ of data, a factor of 6 more than in previous measurements~\cite{bib:prevmeas},
which results in an extension of the $p_T$ coverage by 100~GeV/c. The cross section is measured up
to 400~GeV/c, testing the pQCD over 6 orders of magnitude. Photons are required to have
$|\eta|<$1.0, $p_T>$30~GeV/c, and to be isolated with $E_T^{iso}<$2.0~GeV, where $E_T^{iso}$ is
defined as the transverse energy deposited in a cone of
radius $R=\sqrt{\Delta\eta^2+\Delta\phi^2}=$0.4 around the photon candidate minus that of the
photon. To reject electrons from $W$ decays and non-collision backgrounds, candidates with
$E_T^{miss}>$0.8$p_T$ are vetoed. The fraction
of prompt photons in the sample is determined by fitting the calorimeter isolation distribution in the data to
signal and background templates obtained from MC simulations. Data are unfolded back to hadron level
using a bin-by-bin procedure in a PYTHIA inclusive photon MC sample. The unfolding factors correct for selection
efficiencies and for the detector resolution and acceptance effects, and vary between 64\% to 69\%
in the $p_T$ considered. The systematic uncertainties in the cross section range from 10 to
15\%, dominated by the purity determination at low $p_T$ and by the uncertainty in the photon energy
scale at high $p_T$. \\
The cross section is measured as a function of the photon transverse momentum and compared to NLO pQCD
predictions as given by the JETPHOX~\cite{bib:jetphox} program, with CTEQ6.1M PDFs~\cite{bib:pdf},
BFGII fragmentation functions~\cite{bib:ff} and renormalization, factorization and fragmentation
scales set to the transverse momentum of the photon (see Fig.~\ref{fig:incxsec}). The pQCD predictions have
been corrected to account for the non perturbative effect of the underlying event, estimated from
two different sets of tunes in PYTHIA samples~\cite{bib:ue}. This correction decreases the
theoretical cross section by approximately 9\%, constant in $p_T$. The ratio of the measurement to
the theory as a function of the photon transverse
momentum is also presented in Fig.~\ref{fig:incxsec}. Theory and data agree over the whole measured $p_T$
range except for $p_T<$40~GeV/c. For $p_T>$40~GeV/c the data is over the theory, reproducing the
trend seen in previous measurements.

\section{Photon plus jet cross section}
\label{sec:phojet}
\begin{figure}[tbp]
   \begin{center}
   \begin{tabular}{cc}
        \renewcommand{\tabcolsep}{0.01in}
\includegraphics[width=2.5in, angle=0]{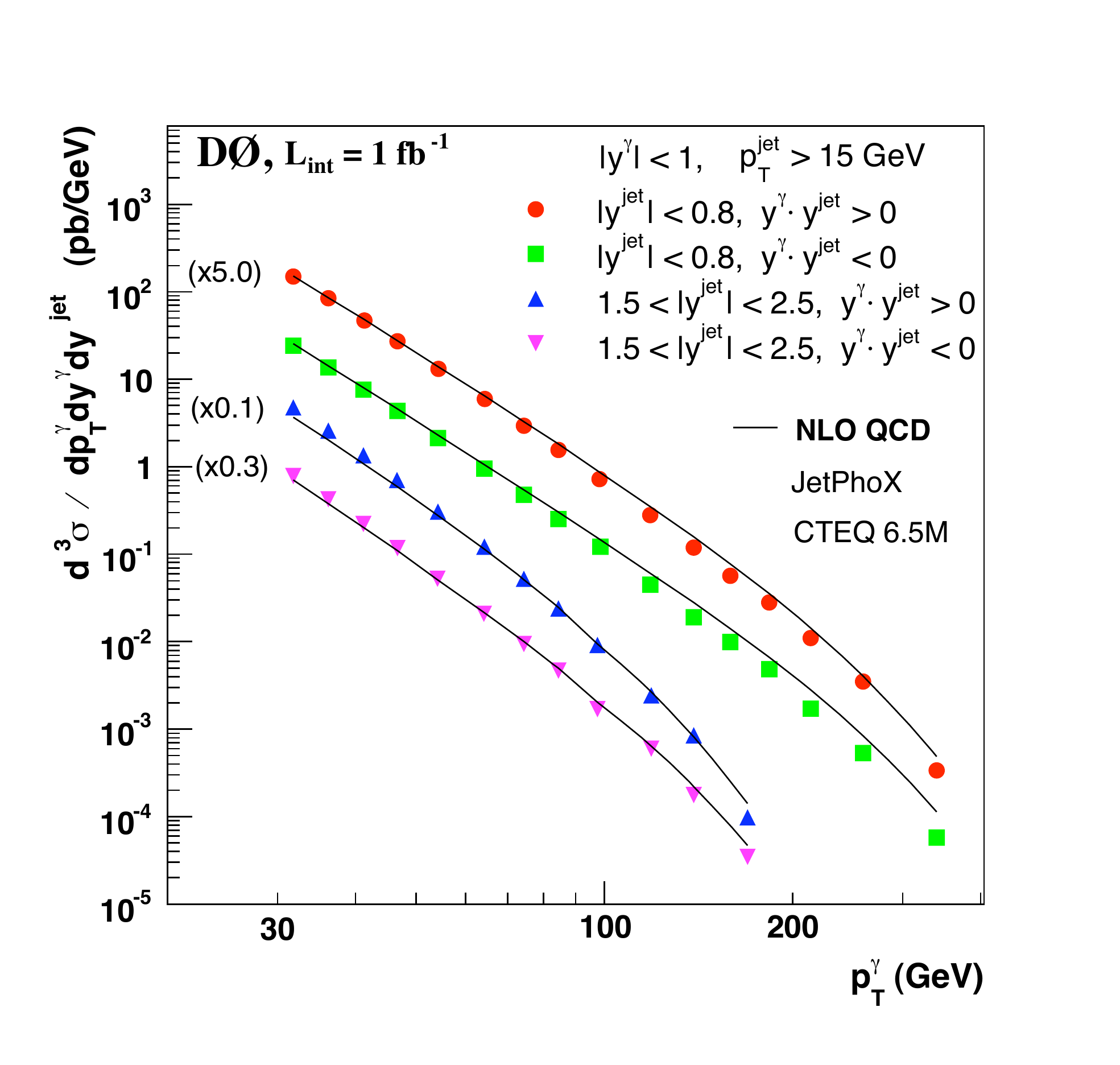} & \includegraphics[width=2.5in, angle=0]{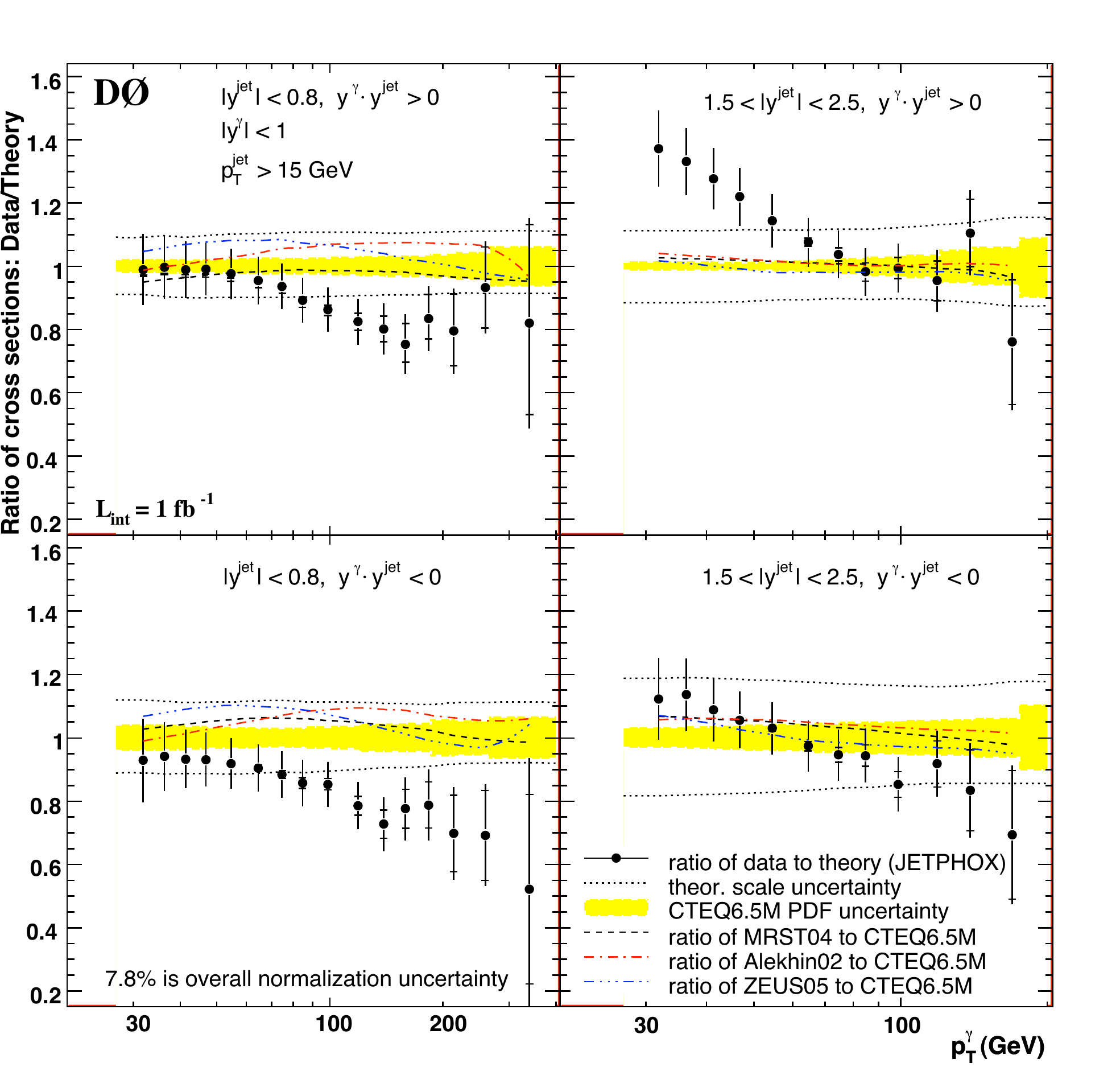} \\
   \end{tabular}
  \end{center}
     \caption{\small The measured cross section as a function of the photon transverse momentum for
the four different kinematic regions (left). For presentation purposes, the cross section result for central
jets and $y^{jet}\cdot y^\gamma>$0 and for forward jets with $y^{jet}\cdot y^\gamma>$0 and
$y^{jet}\cdot y^\gamma<$0 are scaled by factors of 5, 0.1, and 0.3 respectively. The corresponding
ratios to the theory predictions {\it vs} the $p_T$ of the photon, for each measured interval, are shown in the right.
   \label{fig:phojet}}
 \end{figure}

The D\O~Collaboration has measured the photon + jet cross section with 1.1~fb$^{-1}$ of
data~\cite{bib:phojetpaper}. Photons must be isolated and have $|y^{\gamma}|<$1.0 and
$p_T^{\gamma}>$30~GeV/c. The isolation criteria requires the
transverse energy not associated to the photon 
in a cone of radius R = 0.4 around the photon direction to be less than 0.07 times the energy 
of the photon. Backgrounds from cosmics and electrons from W boson decays are vetoed by 
a missing transverse energy requirement of $E_T^{miss} <$ 12.5 GeV + 0.36$p_T^{\gamma}$. Jets are
reconstructed with a midpoint cone algorithm with R~=~0.7, and must have $p_T^{jet}>$15~GeV/c and can be either
central ($|y^{jet}|<$0.8) or forward (1.5$<|y^{jet}|<$2.5). The leading photon and jet are required
to have $\Delta R(\gamma,jet)>$0.7. The photon purity of the sample is determined with a neural
network (NN) using information from the calorimeter and the tracker. \\
The differential cross section as a function of the $p_T$ of the photon is shown in
Fig.~\ref{fig:phojet}. The measurement is performed in four different kinematic regions defined by
central (forward) jets and same  and opposite sign photon and jet
rapidities. Data are compared to theoretical predictions obtained from JETPHOX with CTEQ6.5M PDFs
and BFGII fragmentation
functions. The scales have been chosen to be $\mu_{R,F,f}=p_T^\gamma f(y^*)$ with
$f(y^*)=\sqrt{1+exp(2|y^*|)/2}$ and $y^*=0.5(y^\gamma-y^{jet})$. Non-PQCD effects were considered to
be negligible. As shown in Fig.~\ref{fig:phojet}, the predictions do not describe the shape of
the data for the whole measured range, especially for $|y^{jet}|<$0.8 and $p_T^\gamma>$100~GeV/c and
for 1.5$<|y^{jet}|<$2.5, $y^\gamma\cdot y^{jet}<$0 and $p_T^\gamma<$50~GeV/c, where the difference
in shape is similar to those observed in previous inclusive photon
measurements~\cite{bib:prevmeas}. The comparison to predictions with different
PDFs~\cite{bib:otherpdfs} leads to similar conclusions.

\section{Photon plus heavy flavor jet cross section}
\label{sec:phohfjet}
\begin{figure}[tbp]
   \begin{center}
   \begin{tabular}{cc}
        \renewcommand{\tabcolsep}{0.01in}
\includegraphics[width=2.5in, angle=0]{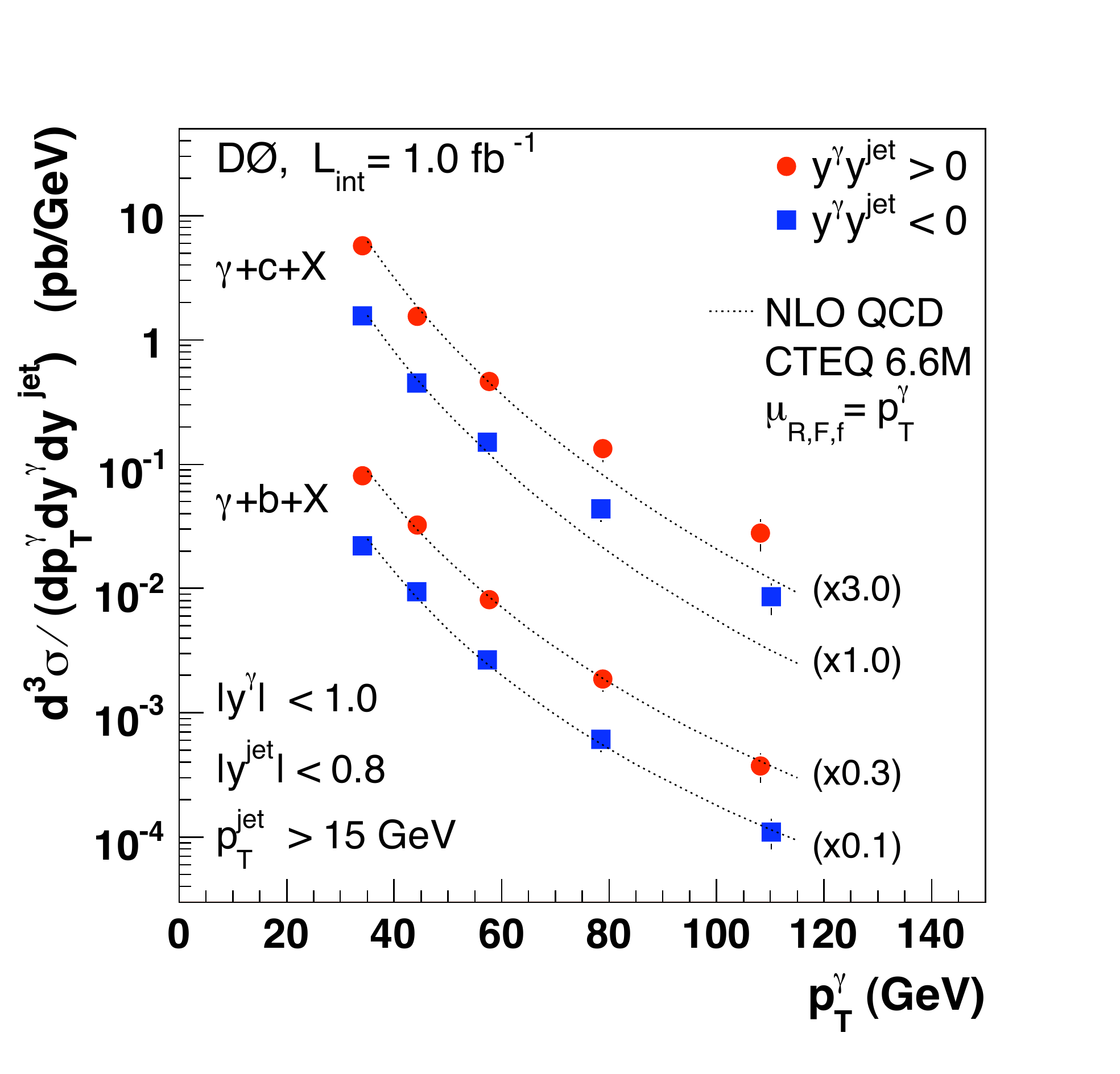} & \includegraphics[width=2.5in, angle=0]{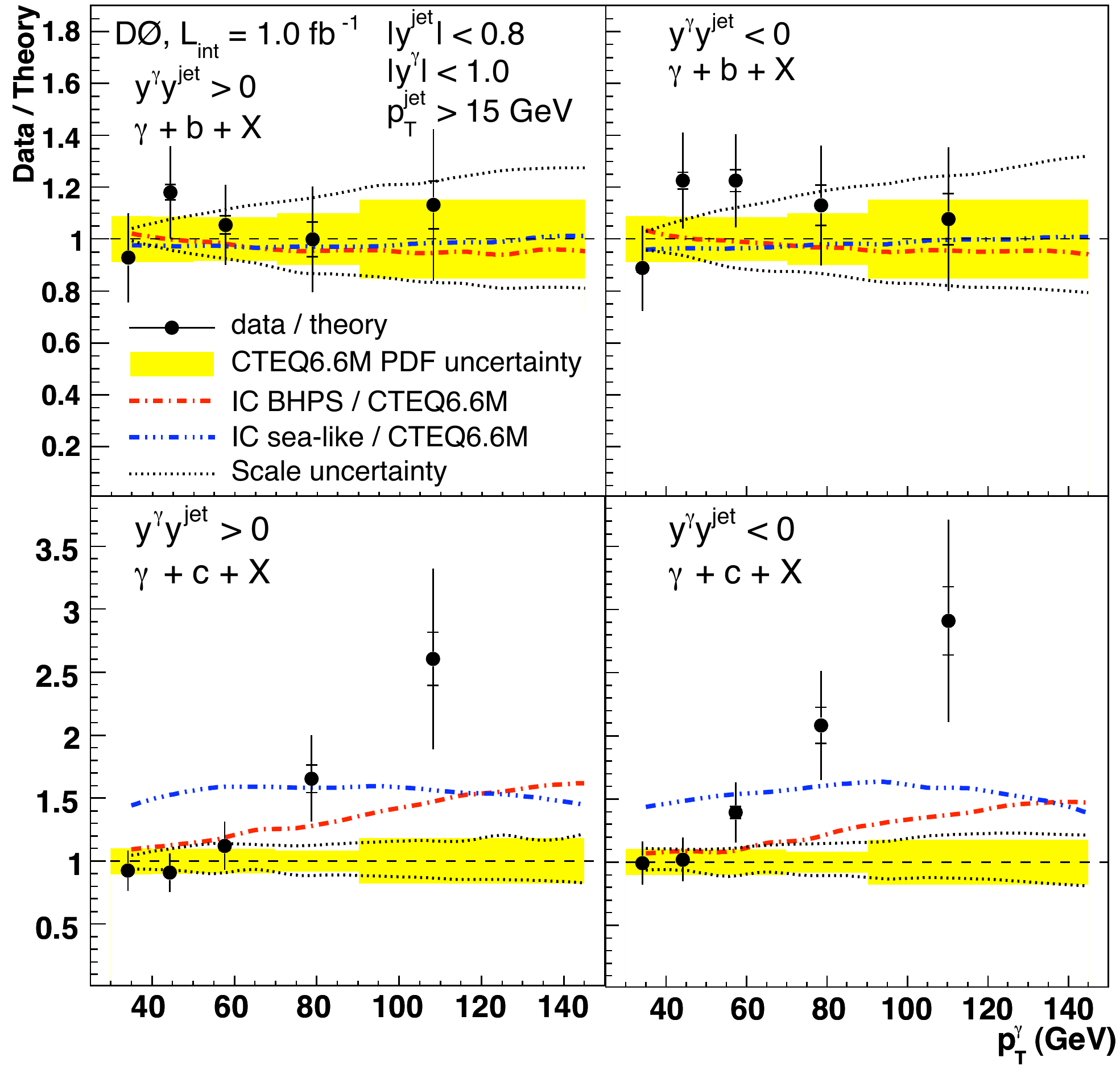} \\
   \end{tabular}
  \end{center}
     \caption{\small The measured $\gamma + b + X$ and $\gamma + b + X$ cross sections as a function
of the photon transverse momentum for the different kinematic regions (left). The ratios to the
theory predictions {\it vs} the $p_T$ of the photon, for each measured cross section, are shown in the right.
   \label{fig:phohf}}
 \end{figure}

The differential cross section for the processes $\gamma + b + X$ and $\gamma + c + X$ has been measured with
1.02~fb$^{-1}$ of data collected by the D\O~detector~\cite{bib:phohfjetpaper}. Photons are subjected to the same selection as
in Section~\ref{sec:phojet}, and jets are reconstructed with a midpoint cone algorithm with
R~=~0.5. The jets are required to have $|y^{jet}|<$0.8 and $p_T^{jet}>$15~GeV/c. The leading jet must have at least two tracks
associated with hits in the silicon microstrip tracker for the heavy flavor tagging. The b and
c jets are identified by using the longer lifetime of the B and D hadrons in a dedicated NN, with an
efficiency of 55-62\% for b jets and of 11-12\% for c jets. The background from light meson decays
is statistically subtracted using a NN. The fractional contribution of b and c jets is determined by
fitting $P_{HF-jet}=-ln\Pi_{i}P^i_{tracks}$ templates to the data, where $P^i_{tracks}$ is the
probability that a track originates from the primary vertex. The templates for b and c jets are obtained from
MC, and the light jet templates come from a data sample enriched with light jets.\\
The cross section is measured as a function of the photon transverse momentum for two different
kinematical regions defined by $y^\gamma\cdot y^{jet}>$0 and $y^\gamma\cdot y^{jet}<$0, and is shown
in Fig.~\ref{fig:phohf} for $\gamma + b + X$ and $\gamma + c + X$. The result is compared to NLO pQCD
predictions with CTEQ6.6M PDFs and $\mu_{R,F,f}=$0.5$p_T^\gamma$. These predictions have been corrected for parton-to-hadron
fragmentation effects by 7.5\% (3\%) in the b (c) jet cross section for low $p_T^\gamma$ and by 1\%
in both predictions at high $p_T^\gamma$. Data and theory agree for the
$\gamma + b + X$ in the whole measured range. For the $\gamma + c + X$ cross section, the theory
does not describe the shape nor the normalization observed in the data for
$p_T^\gamma>$50~GeV/c. Data are also compared to predictions including two models with intrinsic charm
parametrizations in the CTEQ6.6M PDF~\cite{bib:charm}. Both models do not describe the cross
section. The observed difference in the shape could be due to an underestimation of the
$g\rightarrow Q\bar{Q}$ splitting in the theory, which becomes dominant at high $p_T^\gamma$~\cite{bib:gluonsplit}.

\section{Summary}
The cross section for inclusive isolated prompt photons, photon + jet and photon produced in
association to a heavy flavor jet have been measured at the Tevatron, with $\sqrt{s}=$1.96~TeV and
with luminosities that range from 1.02~fb$^{-1}$ to 2.5~fb$^{-1}$. The cross sections are compared
to theoretical predictions, providing a test of the pQCD theory up to 6 orders of magnitude. The
measured cross section agrees with the theory except for $p_T^{\gamma}<$40~GeV/c, where data and
theory shapes differ in a similar manner as that observed in previous measurements. The photon + jet
cross section has been measured for four different kinematical regions, resulting in similar shapes
as in the inclusive measurement. The theory is not able to describe the measurement in the whole
measured range, especially for $|y^{jet}|<$0.8 and $p_T^\gamma>$100~GeV/c and
for 1.5$<|y^{jet}|<$2.5, $y^\gamma\cdot y^{jet}<$0 and $p_T^\gamma<$50~GeV/c. First measurements on $\gamma + b + X$
and $\gamma + c + X$ cross sections have been presented. Good agreement is observed in the case of $b$ jets,
while theory is below the data in the $c$ jet cross section for $p_T^\gamma>$50~GeV/c.

\section*{Acknowledgments}
I would like to thanks to the organization committee for the invitation and for opportunity to participate in this Conference.


\section*{References}
\begin{thebibliography}{99}

\bibitem{bib:prevmeas}
J. Alliti {\it et al.} (UA2 Collaboration), Phys. Lett.{\bf B} 263 544 (1991); \\
V.M. Abazov {\it et al} (D\O~Collaboration), Phys. Rev. Lett.{\bf 87}, 251805 (2001); \\
D. Acosta {\it et al} (CDF Collaboration), Phys. Rev. D{\bf 65}, 112003 (2002). 

\bibitem{bib:jetphox}
P. Aurenche, M. Fontannaz, J.P. Guillet and E. Pilon, JHEP{\bf 0205}, 028 (2002).

\bibitem{bib:pdf}
CTEQ Collaboration, Phys. Rev. D{\bf 51}, 4763-4782 (1995).

\bibitem{bib:ff}
L. Bourhis, M. Fontannaz and J. Ph. Guillet, Eur. Phys. J. C{\bf 2}, 529-537 (1998).

\bibitem{bib:ue}
R. Field, FERMILAB-PUB-06-408-E (2006). 

\bibitem{bib:phojetpaper}
V.M. Abazov {\it et al.} (D\O~Collaboration), Phys. Lett. B{\bf 666}, 2435 (2008).

\bibitem{bib:otherpdfs}
S. Alekin, Phys. Rev. D{\bf 68}, 014002 (2003); \\
A. D. Martin {\it et al.}, Phys. Lett. B{\bf 604}, 61 (2004).

\bibitem{bib:phohfjetpaper}
V.M. Abazov {\it et al.} (D\O~Collaboration), arXiv:hep-ex/0901.0739 (2008).

\bibitem{bib:charm}
J. Pumplin {\it et al.}, Phys. Rev. D{\bf 75}, 054029 (2007). 

\bibitem{bib:gluonsplit}
C. Amsler, Phys. Lett. B{\bf 1} (2008).


\end{thebibliography}
\end{document}